# On-Chip Terahertz Spectroscopy for Dual-Gated van der Waals Heterostructures at Cryogenic Temperatures

Junseok Seo,[1] Zhengguang Lu,[1] Seunghyun Park,[2] Jixiang Yang,[1] Fangzhou Xia,[1,3] Shenyong Ye,[1] Yuxuan Yao,[1] Tonghang Han,[1] Lihan Shi,[1] Kenji Watanabe,[4] Takashi Taniguchi,[5] Amir Yacoby,[2] and Long Ju*,[1]

[1]Department of Physics, Massachusetts Institute of Technology, Cambridge, MA 02139, USA
[2]Department of Physics, Harvard University, Cambridge, MA 02138, USA
[3]Department of Mechanical Engineering, Massachusetts Institute of Technology, Cambridge, MA 02139, USA
[4]Research Center for Electronic and Optical Materials, National Institute for Materials Science, 1-1 Namiki, Tsukuba 305-0044, Japan
[5]Research Center for Materials Nanoarchitectonics, National Institute for Materials Science, 1-1 Namiki, Tsukuba 305-0044, Japan

*Corresponding author. Email: longju@mit.edu

## Abstract

Van der Waals heterostructures have emerged as a versatile platform to study correlated and topological electron physics. Spectroscopy experiments in the THz regime are crucial, since the energy of THz photons matches that of relevant excitations and charge dynamics. However, their micron-size and complex (dual-)gated structures have challenged such measurements. Here, we demonstrate on-chip THz spectroscopy on a dual-gated bilayer graphene device at liquid helium temperature. To avoid unwanted THz absorption by metallic gates, we developed a scheme of operation by combining semiconducting gates and optically controlled gating. This allows us to measure the clean THz response of graphene without being affected by the gates. We observed the THz signatures of electric-field-induced bandgap opening at the charge neutrality. We measured Drude conductivities at varied charge densities and extracted key parameters, including effective masses and scattering rates. This work paves the way for studying novel emergent phenomena in dual-gated two-dimensional materials.

Van der Waals (vdW) heterostructures of two-dimensional (2D) materials have emerged as a powerful platform to study physics of electron correlation and topology.[1,2] Lots of intriguing ground states have been discovered thanks to various tuning knobs available to this material system, such as composition, stacking order, electrostatic gating, and twist angle.[3-23] To understand the underlying physics of these emergent quantum phenomena, information gained through spectroscopy experiments is needed in addition to electrical transport measurements. However, many conventional spectroscopy techniques cannot be directly applied to this new material system mainly due to two challenges. Firstly, most of the systems require a dual-gated structure, where a top gate and encapsulating dielectric prevent surface probes such as scanning tunneling microscopy and photo-emission spectroscopy from exploring the full phase space controlled by the charge density and the gate electric field.[24-27] Secondly, the relevant energy scale of the electron correlation is much smaller than that of conventional strongly correlated materials. For instance, the lattice period of moiré superlattices much larger than atomic bond lengths in crystals results in smaller energy scales for their emergent phenomena. For optical spectroscopy, directly probing excitations in 2D heterostructures requires the application of photons with wavelength (hundreds to thousands of microns) much larger than the typical size of high-quality devices (~10 μm). The Fourier-transformed infrared photocurrent spectroscopy partially overcame these issues,[28-30] but it cannot be directly applied to metallic states due to the large current noise.

On-chip terahertz (THz) spectroscopy has been demonstrated as an effective technique to study low-energy electron dynamics in micron-sized 2D materials (1 THz = 4.14 meV).[31,32] Different from far-field THz spectroscopy, the on-chip configuration confines the THz electric field between a pair of metal transmission lines to efficiently couple to 2D material flakes. Although it has been employed to study monolayer graphene, superconducting NbN and $K_3C_{60}$,[31-34] applying the same technique to dual-gated vdW heterostructures faces significant additional challenges, especially at liquid helium temperature which is often required for revealing the electron correlation effect in vdW heterostructures. Firstly, a metallic top gate may screen the THz electric field and mask the signal from the underlying material of interest. Secondly, although one can potentially reduce the gate disturbance by using a poorly conducting gate layer such as semiconducting transition metal dichalcogenides (TMDs), this scheme has not been demonstrated at liquid helium temperature due to a dilemma: charges in semiconducting TMDs tend to be frozen at liquid helium temperature due to the metal-insulator transition,[35,36] but using a metallic TMD layer could result in DC and THz conductivities similar to or higher than the layer of interest, for example, magic-angle twisted bilayer graphene (see Section S1 in the Supporting Information for detailed discussions).[3,37]

Here, we devise a novel scheme to tame these obstacles by optically activating and deactivating semiconducting TMDs. We show that TMDs can act as a gate when they become photo-conductive, even when they are originally insulating. After the optical illumination to activate the gates is turned off, charges at both TMD gates and a sample layer of interest are maintained. Importantly, the mobility of charges in the gate layers is greatly reduced in the dark environment and their contribution to THz absorption becomes negligible. We demonstrated this scheme by using dual-gated Bernal-stacked bilayer graphene (BLG) as an example, where we measured the dynamic responses of its Fermi liquids.

We fabricated a device (see Section S2 for the procedure) as illustrated in Figure 1a. On a fused silica substrate, the THz emitter, the dual-gated BLG stack, and the THz receiver are dispersed along a pair of transmission lines. The emitter and receiver are photo-conductive switches based on GaAs. THz pulses are launched by femtosecond-laser-excited carriers in the emitter and detected by the receiver in the form of photocurrent. The electric field of THz pulses are recorded as a function of time delay, which can be later transformed into a frequency domain to reveal the spectrum (see Section S3 for the details of THz measurements). When transmitting through the BLG stack, the THz electric field interacts with the sample through its length along the direction of the transmission lines, instead of the bilayer thickness. As a result, the electric field is reduced due to the sample's absorption, and the transmitted THz signal can be used to extract the optical conductivity spectrum of the stack. In our device, the absorption and conductivity are modulated by the charge density and gate electric field by controlling the back and top gate voltages, $V_{BG}$ and $V_{TG}$. In contrast to previous studies, one important new component in our configuration is a white light (WL) illumination that will be used to modulate the electric properties of the semiconducting gate layers.

Figure 1b shows the details of the BLG stack, where two $WS_2$ flakes serve as the gate layers. The THz waveguide is separated from the BLG stack by the hBN dielectrics. Electrodes are deposited for controlling the electrostatic potentials of BLG and gate layers. We used few-layer graphite flakes to bridge the $WS_2$ flakes and the gold electrodes.

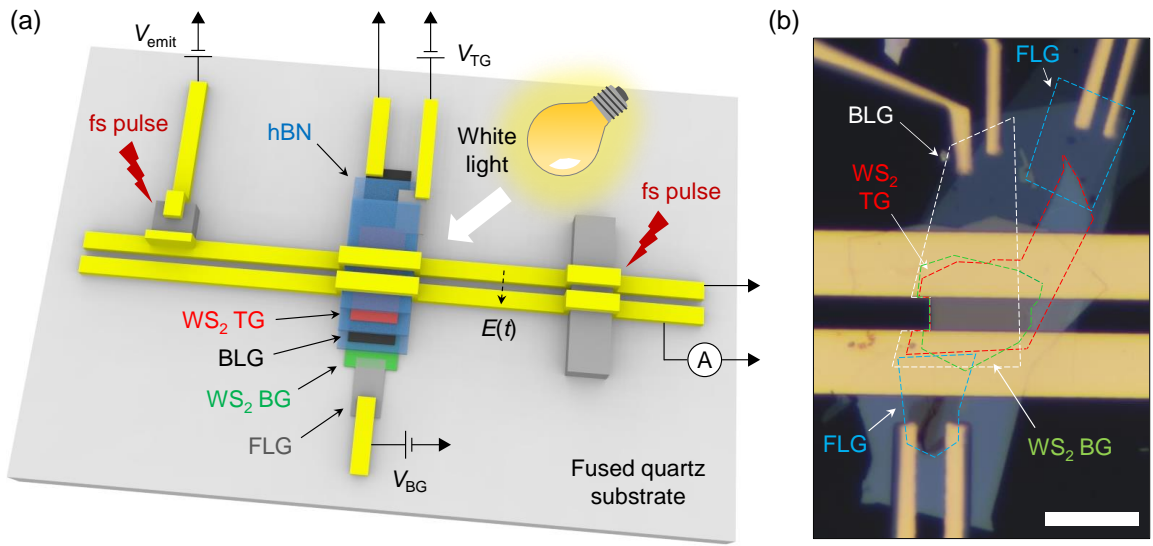

*Figure 1. Device structure and measurement scheme. (a) Schematic of our device and measurement setup from a top view. BG, TG, and FLG denote back gate, top gate, and few-layer graphite, respectively. A current pulse generated in the waveguide by the emitter generates a THz*

*electric field. The transmitted electric field after going through the sample and traveling the waveguide is received by the receiver. White light illumination is used to activate the TMD gates. (**b**) Optical micrograph of our dual-gated BLG device BLG1 embedded in the waveguide. The flakes are outlined by dashed lines with corresponding color. Scale bar, 20 μm.*

The key to a reliable measurement of the BLG THz conductivity is to suppress the free-charge-carrier contribution from the gate layers. This is a dilemma, however, since the mobile charges that supply the gate electric field will unavoidably absorb THz pulses in the static gating scheme and mislead the true absorption by BLG only (see Section S4 for the THz measurements performed with a conducting back gate). We now demonstrate a scheme to solve this problem by combining light illumination, application of gate voltages, and defect/impurity states in $WS_2$. We purposely choose $WS_2$ as the gate, which remains insulating in a wide range of gate voltage at liquid helium temperature.[36] As a result, the gate cannot function in the dark environment due to the lack of free carriers. This is shown as the blue curve in Figure 2a, in which the transmitted photocurrent signal remains constant while the gate voltage is scanned. To activate the gate, we can illuminate it using WL and excite free carriers in $WS_2$. A certain range of WL's photon wavelength is chosen so that it can trigger the photon absorption of $WS_2$ and avoid photo-induced doping effect simultaneously (see Section S3).[38,39] As a result, the transmitted photocurrent signal now responses to the sweeping gate, as shown in the orange curve in Figure 2a. At $V_{BG} = -0.6$ V, the current reaches a maximum, which corresponds to when BLG is in the charge-neutral state. Here, we fix the time delay $t$ at when the photocurrent across the receiver, $I_x(t)$, reaches a positive maximal value, and define this time delay as $t = 0$ as shown in the inset of Figure 2a.

Performing THz measurements with the WL turned on, however, is problematic for several reasons: 1. The photo-excited free carriers in the $WS_2$ gate will absorb the THz pulse and this absorption varies with the gate voltage. Correspondingly, the extraction of the THz response from the BLG will be misinterpreted, and the problem will be even more severe for 2D flat-band systems due to their smaller Drude conductivity; 2. The WL illumination could excite and heat up the BLG, resulting in the measured response being different from the ground state of BLG. To overcome these issues, we developed a procedure that is composed of three steps as shown in Figure 2b and 2c. The left and middle panel in both Figure 2b and 2c illustrate how the $WS_2$ gate and BLG respond to gate voltages when WL is turned off and on, as we have explained above. After the charges settle down in both the $WS_2$ gate and BLG, we can turn off the WL as shown in the third column of Figure 2b and 2c. In the final state with WL off, the photo-excited free carriers in $WS_2$ in the second step get trapped at defect/impurity states. This charge trapping process is different from persistent photoconductivity effect observed in many $SiO_2$-supported TMD phototransistors, where the trapped charges exist at the interface or in the bulk of gate dielectrics.[40] The key point here is that these trapped charges contribute to negligible THz absorption but can keep the gating effect persisting even after turning off the WL. We tested this scheme with other BLG devices (see Section S5 and S6). As a result, we showed the reproducibility of our gating method and confirmed that doped charges can persist for 30 minutes at least. Although the obtained value of persisting time was limited by measurement time, it was still long enough for practical THz measurements.

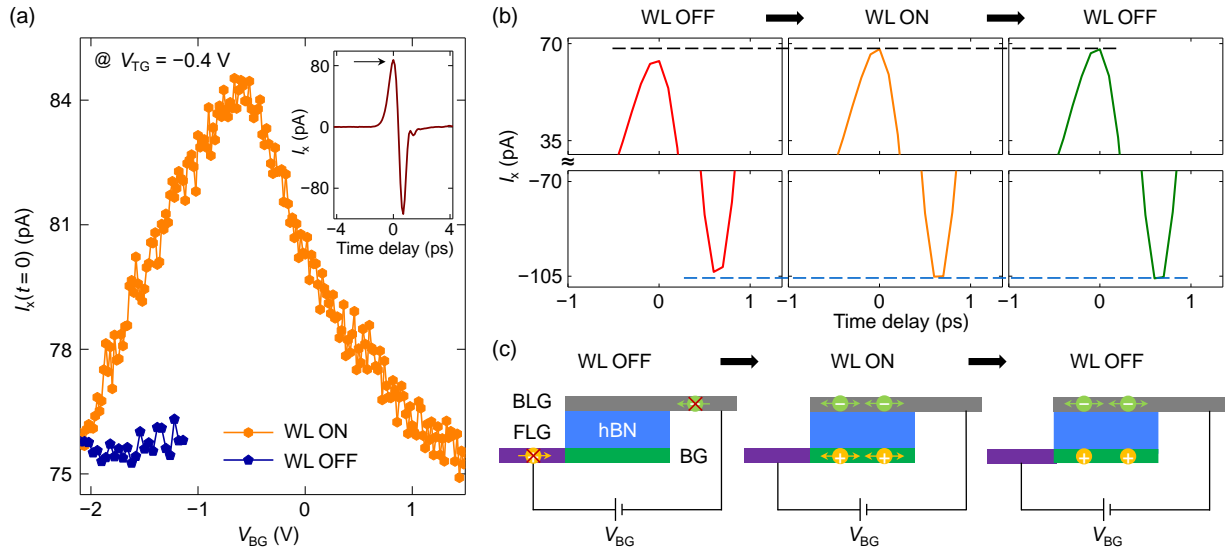

*Figure 2. Optical activation of semiconducting TMD gates at T = 2.5 K. (**a**) Comparison of gating with and without WL. When the WL was turned on, gate voltages were set to be $V_{BG} = -2.1$ V and $V_{TG} = -0.4$ V. Afterwards, $V_{BG}$ was swept to 1.5 V when the WL was turned on (yellow) or off (navy), and $I_x(t = 0)$ was recorded. Inset: Representative time-domain current $I_x$ data. (**b**) Change in a transmitted THz field as the WL is turned on and off again. Gate voltages were set to be $V_{BG} = 2$ V and $V_{TG} = -1$ V when the WL was turned off, and then the time-domain current $I_x$ was recorded (red). Then, the WL was turned on and then $I_x$ was measured (yellow). Subsequently, $I_x$ was measured again after the WL was turned off (green). The maximum positive and negative current were identical for the yellow and green curve (black and blue dotted line, respectively), which showed clear difference from that of the red curve. (**c**) Illustrations of charges in BLG and back gate (BG) layer in the sequence of (**b**). When the WL is turned off (left), charges cannot be injected to the BG and BLG layer since the BG is too insulating. When the WL is turned on (middle), charges are injected and are mobile (represented by double-sided arrows). When the WL is off again (right), charges in the BG layer are frozen and thus cannot contribute to THz absorption.*

Using this gating procedure, we now examine the transmitted THz pulses as a function of charge density $n$ and gate displacement field $D$ which are controlled by the combination of $V_{BG}$ and $V_{TG}$. Before performing the THz measurement at specific $n$ and $D$ to extract the optical conductivity spectrum, we need to understand the basic band structure and experimentally determine the relation between ($V_{BG}$, $V_{TG}$) and ($n$, $D$). Figure 3a shows the current $I_x(t = 0)$ going through the THz receiver as a function of $V_{BG}$ and $V_{TG}$. Here, we tracked the change in the positive current peak, i.e., $I_x(t = 0)$, of the time-domain current data rather than that of the negative peak, due to its more sensitive response to the gate voltages. We observed a bright stripe, which corresponds to increased $I_x(t = 0)$, in the diagonal direction as traced by the red dashed line. In the direction orthogonal to this line, the $I_x(t = 0)$ signal decreases away from the points on the dashed line. Along the dashed line, $I_x(t = 0)$ shows a minimum at the point indicated by the black arrow.

We can understand the $I_x(t = 0)$ map based on the previous studies of the BLG band structure.[41,42] As illustrated in Figure 3b, at $n = D = 0$, BLG is in a charge-neutral state with zero bandgap. In this case, the interband optical transitions result in some THz absorption as represented by vertical arrows in the left panel of Figure 3b. Starting from this scenario, fixing at charge neutrality and increasing $|D|$ will open up a bandgap, which suppresses the interband transition and gives rise to an increase in $I_x(t = 0)$. These expectations agree with the observed $I_x(t = 0)$ signals along the red dashed line in Figure 3a, and the black arrow indicating the $n = D = 0$ point. Fixing $D$ and increasing $|n|$ will induce intraband transitions corresponding to the doped charge carriers, and results in more THz absorption and a decrease in $I_x(t = 0)$. These expectations agree with the observed $I_x(t = 0)$ signals along the orthogonal direction of the red dashed line in Figure 3a.

With this qualitative picture in mind, we try to quantitatively define the relation between ($V_{BG}$, $V_{TG}$) and ($n$, $D$). Figure 3c shows a line-cut along the black dotted line in Figure 3a at varying temperatures. In each curve, the maximal value of $I_x(t = 0)$ corresponds to $n = 0$, i.e., the charge-neutral point (CNP). The determination of $V_{TG}$ corresponding to the CNP becomes more accurate at lower temperatures, as the peak becomes sharper due to reduced thermal fluctuations. For each $V_{BG}$, the corresponding $V_{TG}$ to get CNP can be determined in this way shown in Figure 3c. Repeating this procedure for several $V_{BG}$ values, we can obtain all the CNPs through linear fitting, as shown in Figure 3d. We further determine ($n$, $D$) from ($V_{BG}$, $V_{TG}$) through the analysis of gate capacitances (see Section S7).

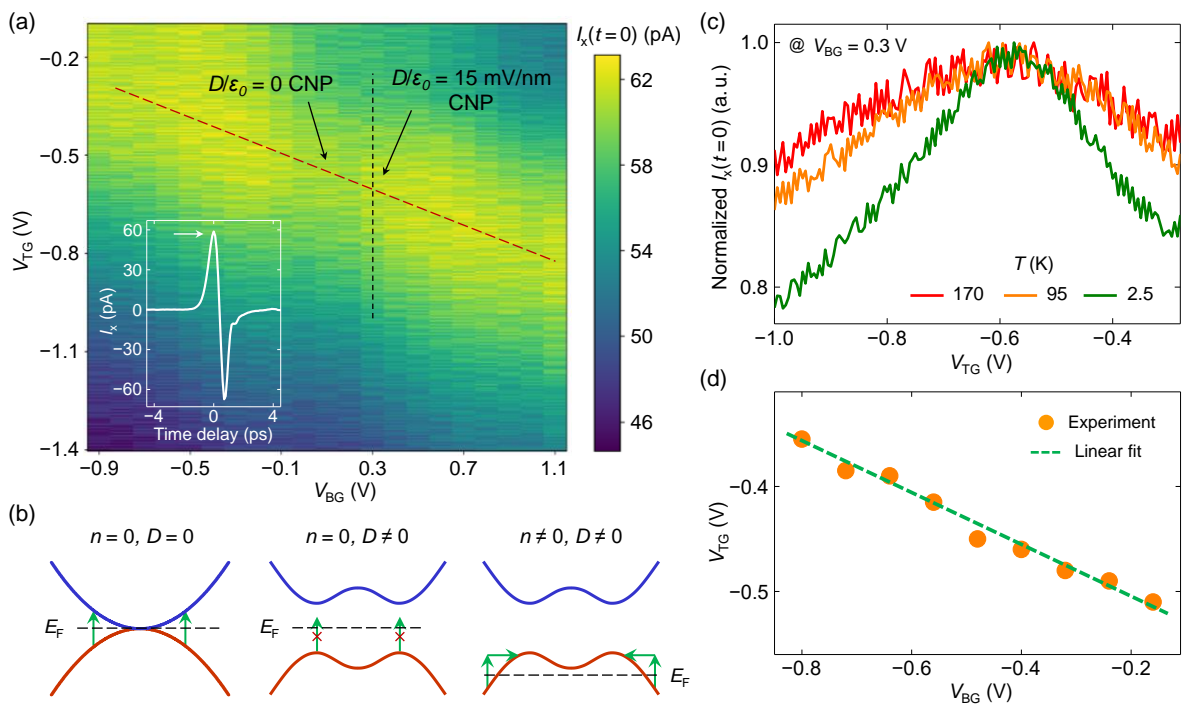

***Figure 3. Optical determination of BLG's CNP.*** (***a***) *Time-domain current $I_x$ at the time delay of zero (white arrow in the inset) as a function of gate voltages. The red dotted line denotes the local maximum of $I_x(t = 0)$ and a point where $I_x(t = 0)$ is minimum along this line is attributed to D = 0 CNP (black arrow).* (***b***) *Schematic illustrations of low-energy BLG band structures that show BLG's THz absorption. When n = 0 and D = 0 (left), some THz pulse can be absorbed by interband transitions. When n = 0 and D ≠ 0 (middle), a bandgap at CNP is open and THz absorption by interband transitions is suppressed. When n ≠ 0 (right), intraband transitions give rise to THz absorption.* (***c***) *Relationship between $I_x(t = 0)$ normalized by its maximum value and $V_{TG}$ for $V_{BG}$ = 0.3 V, taken at three different temperatures. The corresponding line-cut in (**a**) is shown there as a black dotted line.* (***d***) *Relation between $V_{BG}$ and $V_{TG}$ corresponding to maximum $I_x(t = 0)$ from the experimental data (yellow) and its linear fit (green).*

We now turn to the measurements of complex optical conductivity spectra at independently controlled $n$ and $D$. Assuming the THz absorption by $WS_2$ gates can be neglected due to our new scheme of gating as described in Figure 2, the next issue to solve is to find a good reference THz signal to mimic the incident THz pulse totally unaffected by the BLG stack. For far-field THz spectroscopy, the reference signal can be obtained by removing the sample. However, a sample cannot be taken out from our device structure without disturbing the waveguide. Instead, we use the THz signal obtained at $D/\varepsilon_0$ = 80 mV/nm and $n$ = 0 cm$^{-2}$ as the reference, since the bandgap is big enough to suppress all inter-band transitions and the intra-band transitions cannot happen at zero charge density.[28]

We collected the time-domain current $I_x$ at this reference condition and at $n$ = −0.6, −0.9, −1.2 × 10$^{12}$ cm$^{-2}$ and $D/\varepsilon_0$ = 60 mV/nm, as shown in Figure 4a. Similar measurements were done for $n$ = −0.6, −0.9, −1.2 × 10$^{12}$ cm$^{-2}$ and $D/\varepsilon_0$ = 20 mV/nm (Figure 4b). For the fixed $D$, the $I_x(t = 0)$ signal decreases as $n$ increased, suggesting the change is dominated by the intraband transitions.

To extract the spectrum as a function of frequency $f$, we first perform Fourier transformation to obtain both the reference $\epsilon_{\text{ref}}(\omega = 2\pi f)$ and the signal $\epsilon(\omega)$ in the frequency domain. Then, we calculate the optical conductivity as $\sigma(\omega) = \sigma_1(\omega) - i\sigma_2(\omega) = \frac{2W}{Z_0 d}\left[\frac{\epsilon_{\text{ref}}(\omega)}{\epsilon(\omega)} - 1\right]$, which is derived from waveguide electrodynamics.[31] Here, $Z_0$ = 126 Ω is the characteristic impedance of the waveguide obtained from simulations (see Section S8), $W$ = 8 μm is the gap between the two metal lines of the transmission line, and $d$ = 21 μm is the length of BLG along the waveguide direction. This approach compensates for intrinsic and extrinsic effects that may affect the faithful extraction of optical conductivities, such as the frequency-dependent intensity of the generated THz field and the dispersion of a THz pulse during its propagation along the waveguide.

The calculated optical conductivity is shown in Figure 4c and 4d. The data were fitted by the Drude model $\sigma(\omega) = \frac{D^*}{\pi}\frac{1}{\tau^{-1}+i\omega}$, where $D^*$ is Drude weight and $\tau$ is relaxation time. Here, the data are demonstrated up to the frequency of 1.2 THz due to the bandwidth of our THz pulse (see

Section S9). The experimental data follows the model-predicted curves quite well across the whole frequency range, proving the effectiveness of our WL-controlled gating scheme.

The Drude weight is expressed as $D^* = \pi n e^2/m$, where $m$ is the effective mass of charge carriers in BLG. In this regard, we could extract $m$ and $\tau$ at each combination of $n$ and $D$ (see Section S10). The values of $m$ fall into the range of $0.040 – 0.048 m_0$, where $m_0$ is the mass of a bare electron. These are consistent with previous quantum oscillation measurement results.[43-45] We further extract the charge mobility to be of around 56000 cm$^2$ V$^{-1}$ s$^{-1}$, thanks to the simultaneous and independent extractions of $m$ and $\tau$. The obtained mobility supports the value of studying a clean micron-scale hBN-encapsulated sample, rather than exfoliated graphene on SiO$_2$ and large-area graphene grown by chemical vapor deposition.[46,47]

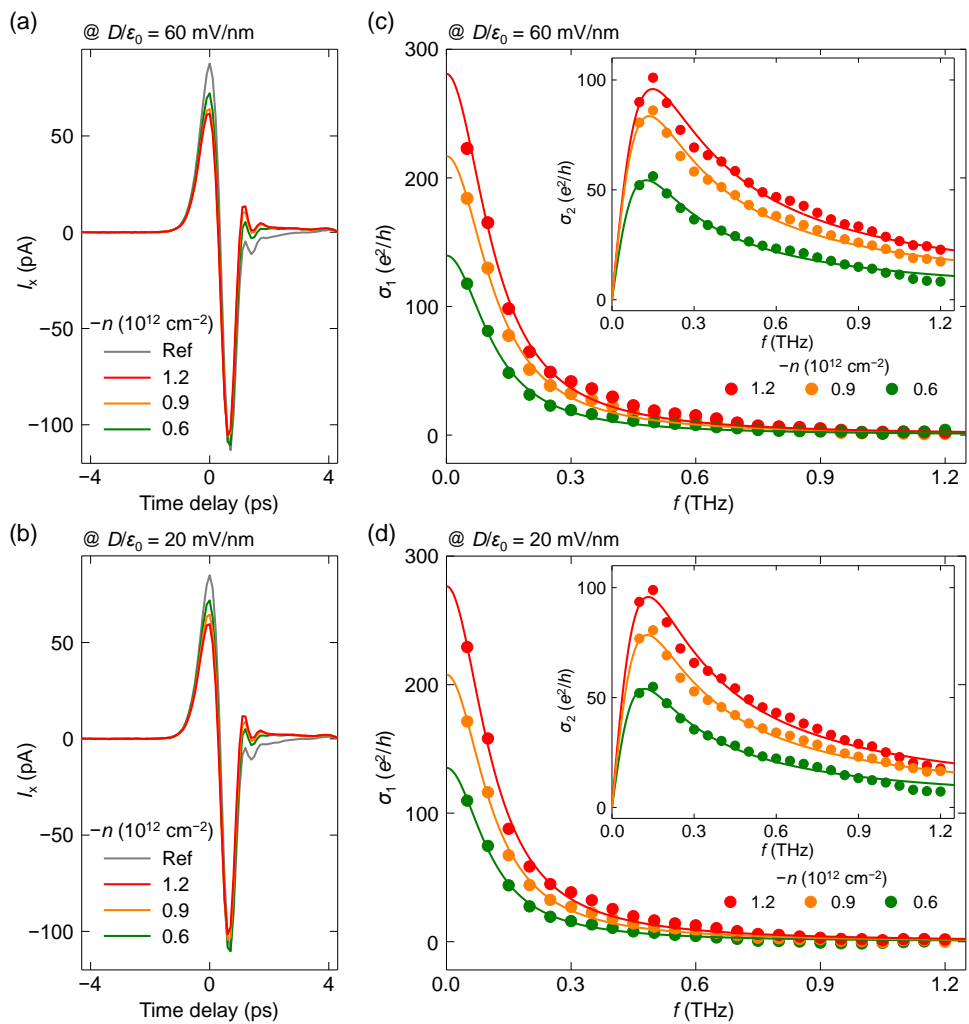

*Figure 4. Optical conductivity spectra of Fermi liquids at T = 2.5 K and varied n and D. Representative time-domain current $I_x$ measured at charge density n = −0.6 (red), −0.9 (yellow), −1.2 (green) × 10$^{12}$ cm$^{−2}$ at (a) D/ε$_0$ = 60 mV/nm and (b) D/ε$_0$ = 20 mV/nm. The reference (Ref, grey) was taken at CNP, D/ε$_0$ = 80 mV/nm. Corresponding spectra of σ$_1$ and σ$_2$ at (c) D/ε$_0$ = 60*

*mV/nm and (**d**) D/ε$_0$ = 20 mV/nm. Dots are the experimental data calculated from (**a**) and (**b**), and lines are fits to the Drude model.*

In summary, we have developed a light-controlled gating scheme for performing the on-chip THz spectroscopy experiment on dual-gated 2D heterostructure devices at and below the liquid helium temperature. This new gating scheme strongly suppresses contribution to THz absorption by the gate layers and allows us to obtain clean optical conductivity spectra in BLG that agree nicely with the Drude model. In this scheme, the carrier density and displacement field could be tuned independently to measure and compare the properties of different ground states of the device. Our work opens up a lot of exciting opportunities for spectroscopic studies on correlated and topological physics in 2D vdW heterostructures, especially in two directions. Firstly, our technique can directly probe charge gaps in the sub-meV to a-few-meV range (see Section 11 for the example of superconducting gap measurements), where the energy gap of many interesting ground states such as superconductors and fractional Chern insulators are expected to reside.[23-25] Secondly, it can be utilized to investigate the charge dynamics of diverse compressible states. THz spectroscopy has been employed to study unconventional metallic states such as non-Fermi liquids,[48] whose signatures have recently been observed in vdW systems as well.[16,49,50] More recently, composite-fermion-like behavior has been observed in a half-filled flat Chern band at zero magnetic field.[20,22] THz spectroscopy could provide valuable insights into such new emergent quantum phenomena. Technically, the on-chip THz spectroscopy covers the right frequency range that corresponds to the relaxation time in graphene and other clean 2D materials. Beyond THz measurements, our gating scheme could be generalized to other advanced spectroscopy and microscopy experiments such as a scanning single-electron transistor,[51] microwave impedance microscopy,[52] and scanning near-field microscopy,[53] which are important techniques for understanding the underlying physics of dual-gated vdW heterostructures.

## Acknowledgements

We acknowledge J. H. Park and J. Kong for sharing their MoS$_2$ grown by chemical vapor deposition. J.S. acknowledges assistance with handling hydrofluoric acid by V. Kamboj and Y. Wang. This work is funded by NSF DMR-2225925. L.J. acknowledges support from the Sloan Fellowship. J.S. acknowledges support from the Jeollanamdo Provincial Scholarship for Study Overseas. K.W. and T.T. acknowledge support from the JSPS KAKENHI (grant nos. 20H00354, 21H05233 and 23H02052) and World Premier International Research Center Initiative (WPI), MEXT, Japan. A.Y. is supported by the Quantum Science Center (QSC), a National Quantum Information Science Research Center of the U.S. Department of Energy (DOE). A.Y. is also partly supported by the Gordon and Betty Moore Foundation through Grant No. GBMF 12762, and by the U.S. Army Research Office (ARO) MURI project under Grant No. W911NF-21-2-0147. This work was carried out in part through the use of MIT.nano's facilities.

## Author contributions

L.J. supervised the project. J.S. built the setup with help from Z.L. and F.X. J.S., J.Y., S.Y. and Y.Y. fabricated the devices with help from T.H. and L.S. S.P. grew NbN films with the supervision of A.Y. J.S. performed the THz measurements. K.W. and T.T. grew the hBN crystals. J.S. and L.J. wrote the paper with input from all authors.

## Competing interests

The authors declare no competing financial interest.

## For Table of Contents Only

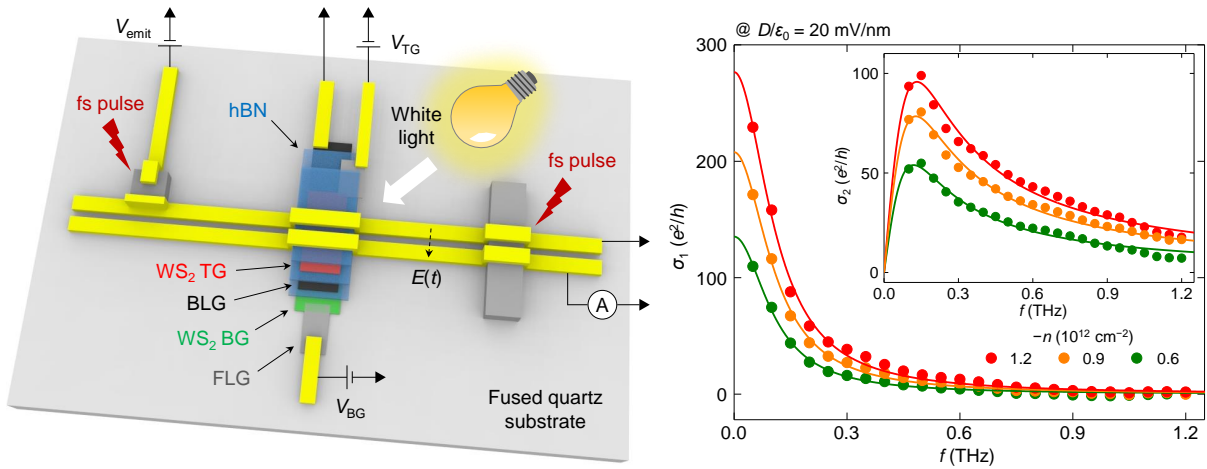

Supporting Information

# On-Chip Terahertz Spectroscopy for Dual-Gated van der Waals Heterostructures at Cryogenic Temperatures

Junseok Seo,[1] Zhengguang Lu,[1] Seunghyun Park,[2] Jixiang Yang,[1] Fangzhou Xia,[1,3] Shenyong Ye,[1] Yuxuan Yao,[1] Tonghang Han,[1] Lihan Shi,[1] Kenji Watanabe,[4] Takashi Taniguchi,[5] Amir Yacoby,[2] and Long Ju*,[1]

[1]Department of Physics, Massachusetts Institute of Technology, Cambridge, MA 02139, USA

[2]Department of Physics, Harvard University, Cambridge, MA 02138, USA

[3]Department of Mechanical Engineering, Massachusetts Institute of Technology, Cambridge, MA 02139, USA

[4]Research Center for Electronic and Optical Materials, National Institute for Materials Science, 1-1 Namiki, Tsukuba 305-0044, Japan

[5]Research Center for Materials Nanoarchitectonics, National Institute for Materials Science, 1-1 Namiki, Tsukuba 305-0044, Japan

Corresponding author's e-mail: longju@mit.edu

1. Estimation of THz absorption by metallic TMDs and flat-band 2D systems

The intensity of THz absorption by a metallic layer doped with the same amount, $n$, of charges can be estimated from its effective mass $m$, since the Drude weight is proportional to $n/m$. Although it is not clear that compressible states in flat-band systems follow the Drude model,[1,2] this can give approximate estimates. The effective mass of TMDs has been studied theoretically through band structure calculations.[3] The effective mass of flat-band 2D systems, such as twisted graphene, has been measured by quantum oscillations.[4-6] Values obtained in literature are shown in Table S1. The effective mass of twisted graphene systems can be comparable or even larger than that of TMDs. This is also reflected in their DC conductivities (i.e., $f = 0$) at low temperature. The resistance of doped metallic TMDs ranges from hundreds of ohms to few tens of kiloohms.[7-10] The longitudinal resistance of flat-band 2D systems is a few to tens of kiloohms,[1,4-6] which is comparable or even higher than that of TMDs.

2. Device fabrication

We started from the preparation of photoconductive switches for THz measurement. A two-inch wafer supplied by BATOP GmbH consisted of 2.6-µm-thick low temperature-grown gallium arsenide (LT-GaAs) with growth temperature of 300 °C, 500-nm-thick $Al_{0.9}Ga_{0.1}As$, and GaAs substrate from the top. The shape of LT-GaAs pieces was determined by an etch mask (LOR 3A and AZ 3312) defined by home-made photolithography. Then, exposed LT-GaAs was etched by a solution of citric acid and hydrogen peroxide, and the etch mask was removed by Remover PG (Kayaku Advanced Materials). Subsequently, $Al_{0.9}Ga_{0.1}As$ was dissolved by 10% hydrofluoric acid. As a result, etched LT-GaAs pieces were loosely attached to the GaAs substrate. Polypropylene carbonate (PPC)/polydimethylsiloxane was used to transfer the etched pieces to target positions on a fused quartz substrate (MSE Supplies). PPC remaining on the substrate was removed by annealing in high vacuum (Figure S1a).

Next, individual flakes were mechanically exfoliated onto $SiO_2$/Si substrates from bulk crystals of graphite (NGS, graphite.de), hBN, $WS_2$ (HQ Graphene), and $MoS_2$ (SPI Supplies). The thickness of hBN flakes was measured by atomic force microscope (Innova, Bruker). For BLG1 (Figure 1b), the bottom stack composed of hBN, FLG, and few-layer $WS_2$ was assembled by a dry transfer technique with poly(bisphenol A carbonate)/polydimethylsiloxane. After the stack was released onto the prepared substrate, remaining polymer was removed in chloroform and the top hBN surface was cleaned by an atomic force microscope tip (Figure S1b). A remaining stack comprising of hBN, FLG, few-layer $WS_2$, hBN, and BLG was assembled and transferred onto the bottom stack in the same way (Figure S1c). For BLG2 (Figure S4a) and BLG3 (Figure S5a), few-layer WS$_2$ in the bottom stack was substituted with few-layer MoS$_2$ while maintaining the other structures.

Electrical connection to FLG and BLG was made by photolithography (MLA 150, Heidelberg), reactive ion etching (Ekrub Nano), and thermal evaporation of chromium and gold (DTT, Element π). The THz waveguide (5 nm titanium and 220 nm gold) was formed by photolithography (MLA 150, Heidelberg) and electron-beam evaporation (ATC-2036, AJA) (Figure S1d). Possible native oxide on LT-GaAs surface was removed by hydrochloric acid.[11]

3. THz measurement

The beam path of our THz measurement setup is shown in Figure S2. Laser pulses to illuminate LT-GaAs photoconductive switches came from femtosecond laser (Origami O-08LP, NKT Photonics). The laser output passed a long pass filter (FEL0650, Thorlabs) to block unwanted frequency components. The laser pulse to illuminate a detector passed a delay stage (NRT100, Thorlabs), which was used to control a time delay $t$ between the other laser pulse heading to an emitter that passed a stage equipped with a piezoelectric shaker (DRV157, Thorlabs). The piezoelectric shaker provided a sinusoidal unidirectional vibration with amplitude of ~13 μm and frequency of $f_P$ = 27.111 Hz to add an AC modulation to the time delay $t$ in the lab time as $t(T) = \langle t \rangle + A\sin(2\pi f_P T)$. Here, $A$ is the amplitude of the modulation, $T$ is the lab time, and $\langle t \rangle$ is the lab-time-averaged time delay. Both laser pulses were focused to the photoconductive switches with a beam spot size of ~10 μm. Typical power of the pulse illuminating the emitter (detector) was 5 mW (3 mW). An electrical current across the detector was inverted into a voltage signal by a current preamplifier (SA-606F2, NF Corporation) and then was fed into a lock-in amplifier (SR7225, Signal Recovery) with reference given by the piezoelectric shaker.

For the collection of time-domain $I_x$ data throughout this work, a time delay step of 0.1 ps was used. The AC modulation given by the piezoelectric shaker results in the modulation of the THz electric field $E(t)$ in the lab time as $E[t(T)] = E[\langle t \rangle + A\sin(2\pi f_P T)] \approx E(\langle t \rangle) + \frac{dE}{dt}\big|_{\langle t \rangle} A\sin(2\pi f_P T)$, for the small enough modulation amplitude. Since the photocurrent across the detector locked-into the shaker's reference is proportional to the electric field, $I_x$ effectively measures $dE/dt$. Each trace of $I_x(t)$ was measured up to three times for BLG1 device, due to its large enough THz absorption, and then averaged out. For the averaging process, we performed a scan of $\varepsilon_{ref}(\omega)$ right after taking a single scan of $\varepsilon(\omega)$, and then averaged out $\varepsilon_{ref}(\omega)/\varepsilon(\omega)$. Compared to a procedure to average out $\varepsilon_{ref}(\omega)$ and $\varepsilon(\omega)$ individually, this way has advantage to minimize the long-term drift of femtosecond laser alignment.

Optical illumination to activate TMD gates was given by warm white light (MWWHL4, Thorlabs). The white light passed two long pass filters (FGL530M and FGL550M, Thorlabs) to avoid photoinduced doping by defect states in hBN.[12] We found using stronger white light makes the TMD gates more photo-conductive and thus leads to a faster charging process. Then, the white light illuminated the sample globally when charging BLG with TMD gates and was turned off after the charging process was completed to get rid of possible heating effect.

Samples were mounted in a continuous flow cryostat (ST-500, Janis) with a fused silica optical window and high vacuum. Liquid helium was pumped during the measurements, which offered a base temperature of 2.5 K. $V_{bias}$ was applied by a DC voltage source (DC205, Stanford Research Systems). $V_{BG}$ and $V_{TG}$ were applied by a source measure unit (GS820, Yokogawa).

4. Optical conductivity spectra of graphene gated by a conducting $MoS_2$ back gate

Figure S3a displays a monolayer graphene device back-gated by $MoS_2$ embedded in THz waveguide. In contrast to exfoliated $MoS_2$, this was grown by chemical vapor deposition (CVD) and was much more conducting in the dark environment, confirmed by its gating range at the base temperature in the absence of WL illumination. When the graphene is electron-doped (hole-doped), the $MoS_2$ gate is hole-doped (electron-doped). Due to the unipolar transport characteristics of $MoS_2$,[7] its THz absorption becomes more and more significant as the graphene is hole-doped, that is, negative gate voltage $V_G$ is applied. Figure S3b shows the optical conductivity spectra when the graphene was electron-doped. Here, the reference was chosen to be $I_x(t)$ taken at the CNP, following the previous measurements.[13] The data obeyed the Drude model reasonably well with a relaxation time much shorter than the data for BLG1, because of lower sample quality from bubbles generated during the transfer of CVD-grown $MoS_2$. Figure S3c demonstrates the spectra when more negative value of $V_G$ was applied. These spectra do not follow the simple Drude model more considerably as $V_G$ decreases. We attribute this to the conducting $MoS_2$ gate that can disturb graphene's THz absorption. This stresses the importance of being able to ignore the THz absorption of a gate layer.

5. Optical activation of TMD gates in the device BLG2 probed with electrical transport

We present data from another device that demonstrates our scheme to activate TMD gates. Figure S4a displays the image of a dual-gated BLG device under electrical transport measurements in a DC limit. Figure S4b shows an electrical current $I_{23}$ as a function of $V_{BG}$ when voltage of 88 μV was applied across the two current leads. When the white light was turned off, $I_{23}$ stopped changing from $V_{BG} \approx 0$ V, which indicates the back gate became too insulating to serve as a gate. In stark contrast, $I_{23}$ shows a highly symmetric shape around $V_{BG} = -0.55$ V and keeps increasing up to 8 V, when the white light was turned on and the back gate was activated. We can exclude the possibility of photoinduced doping by the photoexcitation of either interfacial[14] or bulk[12] charge traps from hBN from the observation that $I_{23}$ follows the same curve from −6 to −1.5 V where the back gate was still working properly without optical illumination. To examine if charges doped by optically assisted gating maintain after the white light is turned off, we monitored the current $I_{23}$ and voltage difference $V_{41}$ as a function of time before and after turning off the light. The data at three different $V_{BG}$ values are shown in Figure S4c. Both $I_{23}$ and $V_{41}$ did not change after turning off the light. This test has been performed for 30 minutes (Figure S4d) and we could not find evidence for doped charges being drained out. Since it takes around two minutes to obtain one curve of time-domain current, we conclude doping by TMD gates stays stable enough during the THz measurements.

6. Data from the device BLG3

Figure S5a shows the image of a dual-gated BLG device BLG3. The width of the waveguide was ≈8.1 μm and the length of BLG along the waveguide direction was ≈7.4 μm. Figure S5b displays the side-view schematic of the vdW heterostructure embedded in the THz waveguide. We measured the time-domain current $I_x(t)$ at one configuration of the gate voltages, fixed the time delay at zero, and recorded $I_x(t=0)$ as a function of $V_{BG}$ and $V_{TG}$ (Figure S5c). Along with a purple dotted line corresponding to the CNP of BLG, the other line of increased $I_x(t=0)$ (white dotted line) with the slope same to the CNP line's appears at more negative values of $V_{TG}$. We attributed this to a feature from moiré full filling (i.e., filling factor of −4), where a moiré superlattice arises from the BLG and top hBN. From this, we could estimate the relative angle between the BLG and top hBN to be roughly 0.7°. In addition, the slope of both the purple and white dotted line was ≈−0.76, which is in good agreement with a ratio between the bottom and top hBN thickness (18.5/25 = 0.74) determined by atomic force microscopy. In Figure S5d, we tested our gating scheme to this device. First, the WL was turned off and the gate voltages were set to $V_{BG} = 2$ V and $V_{TG} = -1$ V. Then, the current $I_x$ was measured (Figure S5d, left). Next, $I_x$ was measured again after the WL was shone (Figure S5d, middle). Finally, the WL illumination was turned off and $I_x$ was measured (Figure S5d, right). The observed behavior is identical to what has been discussed in Figure 2b and 2c in the main text.

7. Determination of $(n, D)$ from $(V_{BG}, V_{TG})$ from the CNP line for the device BLG1

We first discuss if our optical method to determine BLG's CNP is reasonable. From the linear fit in Figure 3d in the main text, we can determine the slope of the CNP line to be ≈−0.25. This value is consistent with a ratio between bottom and top hBN thickness 13/47 ≈ 0.28) determined by atomic force microscopy. This supports the experimental validity of our method. A similar method targeting the maximum transmitted THz field has been demonstrated for determining the CNP of monolayer graphene.[13,15]

We now discuss the determination of a relationship between ($n$, $D$) and ($V_{BG}$, $V_{TG}$). This relation should be determined individually for each device, because of possible residual charges that may vary from sample to sample. For the device BLG1, we can attribute this offset for the back and top gate voltage, $V_{BG,0}$ and $V_{TG,0}$, respectively, to the gate voltages corresponding to the $D = 0$ CNP (black arrow in Figure 3a). Consequently, we calculate the carrier density and displacement field as $n = \frac{\varepsilon_0 \varepsilon_r}{-e}\left(\frac{V_{BG}-V_{BG,0}}{d_B} + \frac{V_{TG}-V_{TG,0}}{d_T}\right)$ and $D = \frac{\varepsilon_0 \varepsilon_r}{2}\left(\frac{V_{BG}-V_{BG,0}}{d_B} - \frac{V_{TG}-V_{TG,0}}{d_T}\right)$. Here, $\varepsilon_0$ is vacuum permittivity, $\varepsilon_r = 3.6$ is the dielectric constant of hBN, $e$ is electron's charge, $d_B = 47$ nm ($d_T = 13$ nm) is the thickness of bottom (top) hBN gate dielectrics, and $V_{BG,0} = 0.1$ V and $V_{TG,0} = -0.56$ V are the offset of the gate voltages.

## 8. Simulation of a hollow waveguide

We used 2D mode solver in COMSOL Multiphysics to simulate our waveguide and extract its characteristic impedance $Z_0$. Figure S6a displays a 2D color plot of the electric field $E$ at fixed frequency of 0.1 THz. For the gold waveguide, we used the resistivity of gold, $\rho = 7.8$ nΩ·m, obtained from a four-probe measurement for a gold Hall bar evaporated by the same equipment. For fused quartz, we put its relative permittivity of 3.85 obtained in literature.[16] We found there is a large electric field $E_y$ perpendicular to the plane of a sample near the edge of the waveguide. However, we expect the horizontal component $E_x$ dominates the THz absorption by the sample, due to the anisotropy in the electrical conductivity of graphene systems and the ultrathin nature of 2D materials,[17,18] resulting in a negligibly small dipole moment in the $y$ direction.

To calculate $Z_0$ corresponding to the waveguide, we should extract potential difference $V$ between the two metal lines when an electric current $I$ flows along the waveguide. These can be calculated as $V = -\int_{\text{Path 1}} \vec{E} \cdot d\vec{l}$ and $I = \oint_{\text{Path 2}} \vec{H} \cdot d\vec{l}$ by Ampere's law, where $\vec{H}$ is a magnetic $H$ field and the path 1 and 2 are described in Figure S6b. Accordingly, $Z_0 = V/I$. Figure S6c shows refractive index $n$ and impedance $Z_0$ calculated at several values of frequency $f$. We could extract $n = 1.56$ and $Z_0 = 126$ Ω, which are mostly independent of the frequency. Experimentally, we could estimate $n \approx 1.63$ from THz pulses reflected from the end of a hollow waveguide, which is in good agreement with the calculated value. The extracted value of $Z_0$ was used to calculate optical conductivity.

9. Bandwidth of the generated THz pulse

We here investigate a hollow waveguide (i.e., THz waveguide without any sample embedded) to obtain the bandwidth of our THz pulse. Figure S7a shows the image of our typical hollow waveguide. Its time-domain current $I_x(t)$ is displayed in Figure S7b. Since $I_x(t)$ effectively measures the time derivative of the generated THz field $E(t)$, we can reconstruct $E(t)$ by the numerical integration of $I_x(t)$. The integrated data is shown in Figure S7c. The magnitude of $E(f)$ (i.e., field intensity), where $E(f)$ is the fast Fourier transform of $E(t)$, is demonstrated in Figure S7d. It exhibits a peak at $f = 0$ THz and decreases from its maximum value by around ten times at $f = 1.0$ THz (red dotted line), where the intensity is larger than the top of a noise floor (yellow rectangle) by ~100 times. The field intensity decreases faster at higher frequency, and it becomes larger than the top noise floor by ~50 times at $f \sim 1.2$ THz. From our experience, this signal-to-noise ratio value was marginal to get reliable data of optical conductivity. The improvement of our THz bandwidth is possible by both decreasing the noise and increasing the field intensity at a high frequency region. In particular, the latter may be achieved by incorporating LT-GaAs grown at lower temperature to decrease its carrier lifetime and femtosecond laser with a shorter pulse duration, which is beyond the scope of this paper.

10. Parameters extracted from fitting in Figure 4c and 4d

Data in Figure 4c and 4d could be fitted to the Drude model appreciably well. This also indicates no signature of charges trapped in TMDs was observed, possibly due to energy mismatch between our THz photons (up to 5 to 6 meV) and the energy level of trap states in TMDs.[19,20] Two parameters, effective mass $m$ and relaxation time $\tau$, could be extracted when fitting the obtained $\sigma_1$ to the Drude model (Table S2). The charge mobility could be calculated as $\mu = e\tau/m$.

11. On-chip THz spectroscopy on a superconducting NbN film

Figure S8a shows the image of a 30-nm-thick NbN sample. The NbN film is capped with hBN and embedded under THz waveguide. The sample turns into a superconductor at critical temperature $T_c$ of ~8.4 K (Figure S8b). Figure S8c shows the optical conductivity $\sigma_1$ spectrum of NbN measured at $T = 5.9$ K. Unlike BLG, we could not make NbN insulating to get a reliable THz reference. Alternatively, we measured the time-domain $I_x$ data right above $T_c$ and utilized it as the reference, since NbN usually shows nearly flat $\sigma_1$ up to a few THz.[21,22] This choice underestimates the intensity of a reference pulse, and thus the values of $\sigma_1$ calculated by the equation $\sigma(\omega) = \frac{2W}{Z_0 d}\left[\frac{\epsilon_{\text{ref}}(\omega)}{\epsilon(\omega)} - 1\right]$ become negative at some frequencies. However, the energy corresponding to the dip of $\sigma_1$ is trustable. The spectrum shows an energy gap at $2\Delta \approx 2.9$ meV, which is qualitatively similar to a previous report.[23] Moreover, the temperature dependence of the measured energy gap is consistent with BCS theory, $\Delta(T) = \Delta(0)\tanh\left[\frac{\pi k_B T_c}{\Delta(0)}\sqrt{\frac{T_c}{T}-1}\right]$ (Figure S8d), which reveals that the nature of this gap is a superconducting gap. We also extracted the superfluid stiffness, based on the frequency dependence of THz conductivity $\sigma_2(\omega) = \frac{n_s e^2}{m\omega}$ for photon energy well below $2\Delta$, where $n_s$ is the density of superconducting electrons, $e$ is elementary charge, and $m$ is effective mass.[24] The result shown in Figure S8e is consistent with a phenomenological two-fluid model, $n_s(T) = n_s(0)(1 - T^4/T_c^4)$.[24]

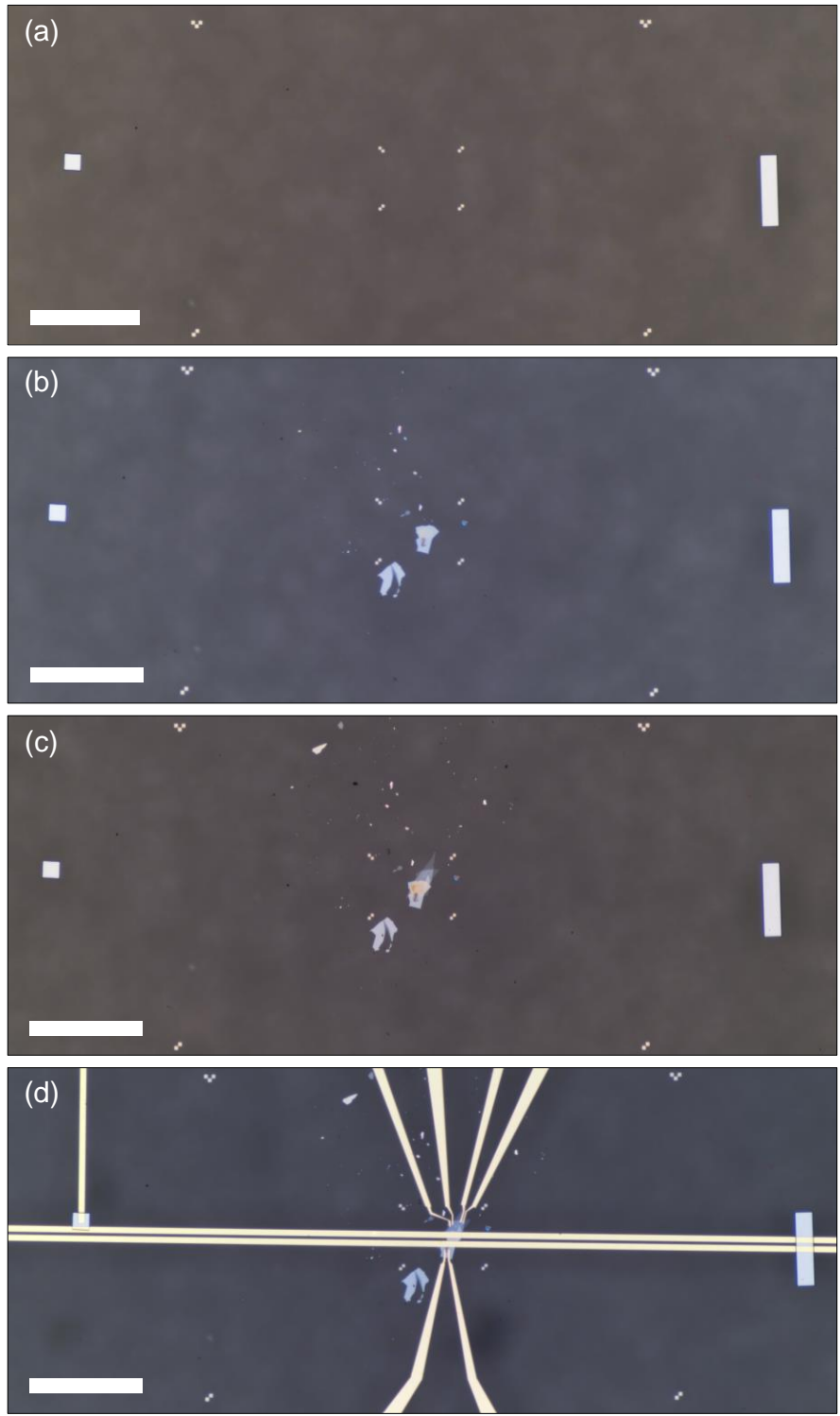

**Figure S1. Sequence of device fabrication.** All scale bars, 0.3 mm. (a) Transfer of LT-GaAs. (b) Transfer of a bottom stack. (c) Transfer of a remaining stack. (d) Electrical connection to top gate, BLG, and back gate, and waveguide fabrication.

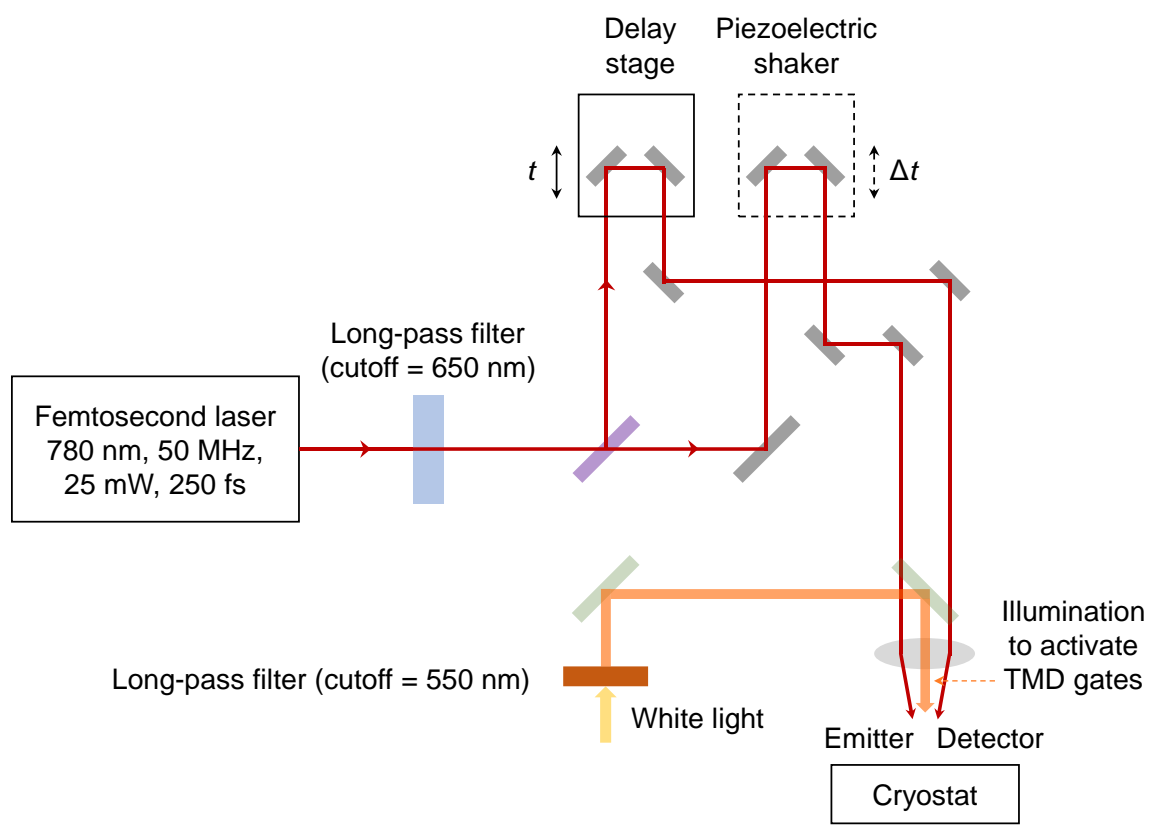

**Figure S2. Block optics diagram of our measurement setup.**

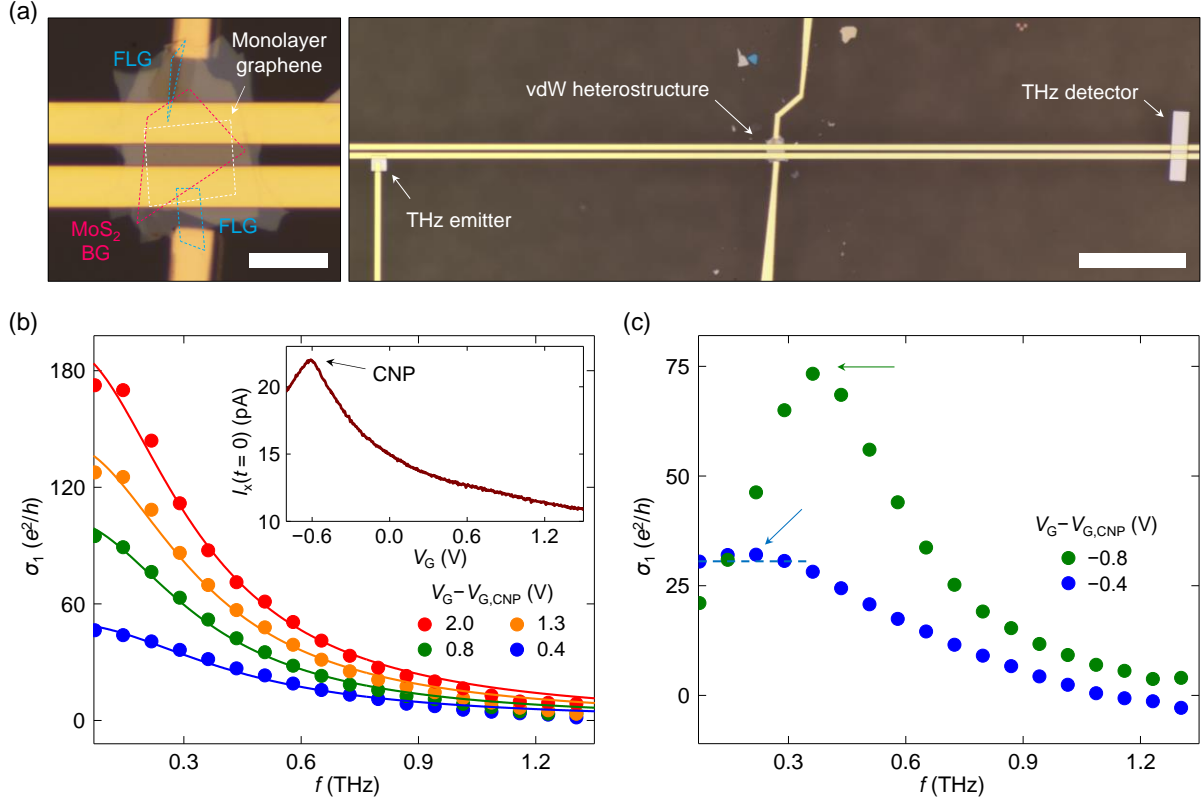

**Figure S3. THz spectroscopy on monolayer graphene with CVD-grown MoS$_2$ back gate conducting in the dark environment.** (a) Optical micrograph of the device. The left panel is the expanded view of a vdW heterostructure in the right panel. Scale bar, 30 μm (left) and 0.3 mm (right). (b) The real optical conductivity spectra of electron-doped monolayer graphene taken at different values of gate voltage $V_G$. Dots are the experimental data, and lines are fits to the Drude model. Inset: $I_x(t = 0)$ as a function of $V_G$. The maximum of $I_x(t = 0)$ corresponding to the CNP was found at $V_G = -0.62$ V. (c) The real optical conductivity spectra of hole-doped monolayer graphene taken at different values of $V_G$. In contrast to the data in (b), maximum $\sigma_1$ indicated as arrows was not located at the smallest frequency, and the whole data deviated from the Drude model.

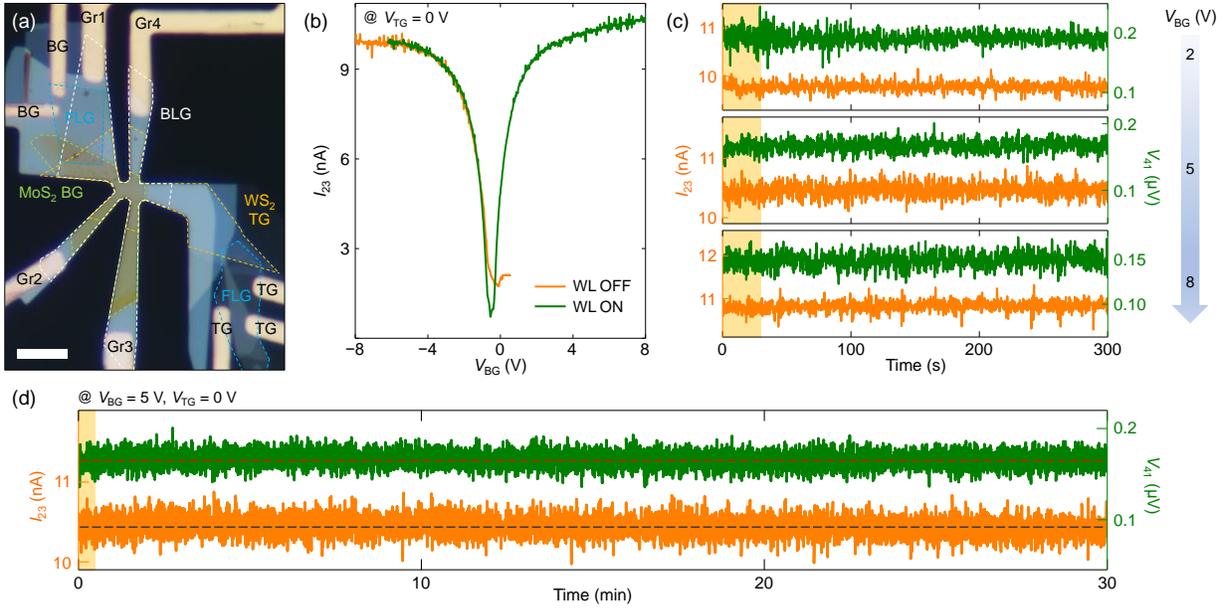

**Figure S4. Electrical transport data of BLG with optically activated TMD gates.** (a) Optical micrograph of the device BLG2. The thickness of bottom and top hBN dielectric was 26 and 17 nm, respectively. Scale bar, 10 μm. (b) Electrical current $I_{23}$ measured at $V_{TG} = 0$ V as a function of $V_{BG}$ when the white light (WL) was turned off (yellow) and on (green). The measurements were conducted by a standard lock-in technique with constant voltage amplitude of 88 μV and frequency of ~41 Hz. $V_{BG}$ was swept from a negative value. (c) $I_{23}$ and $V_{41}$ versus laboratory time at $V_{BG} = 2$, 5, 8 V (top, middle, bottom) and $V_{TG} = 0$ V. Transparent yellow boxes indicate when the white light stayed to be turned on during the measurements. (d) Measurement data same to (c) taken at $V_{BG} = 5$ V and $V_{TG} = 0$ V but for a longer time.

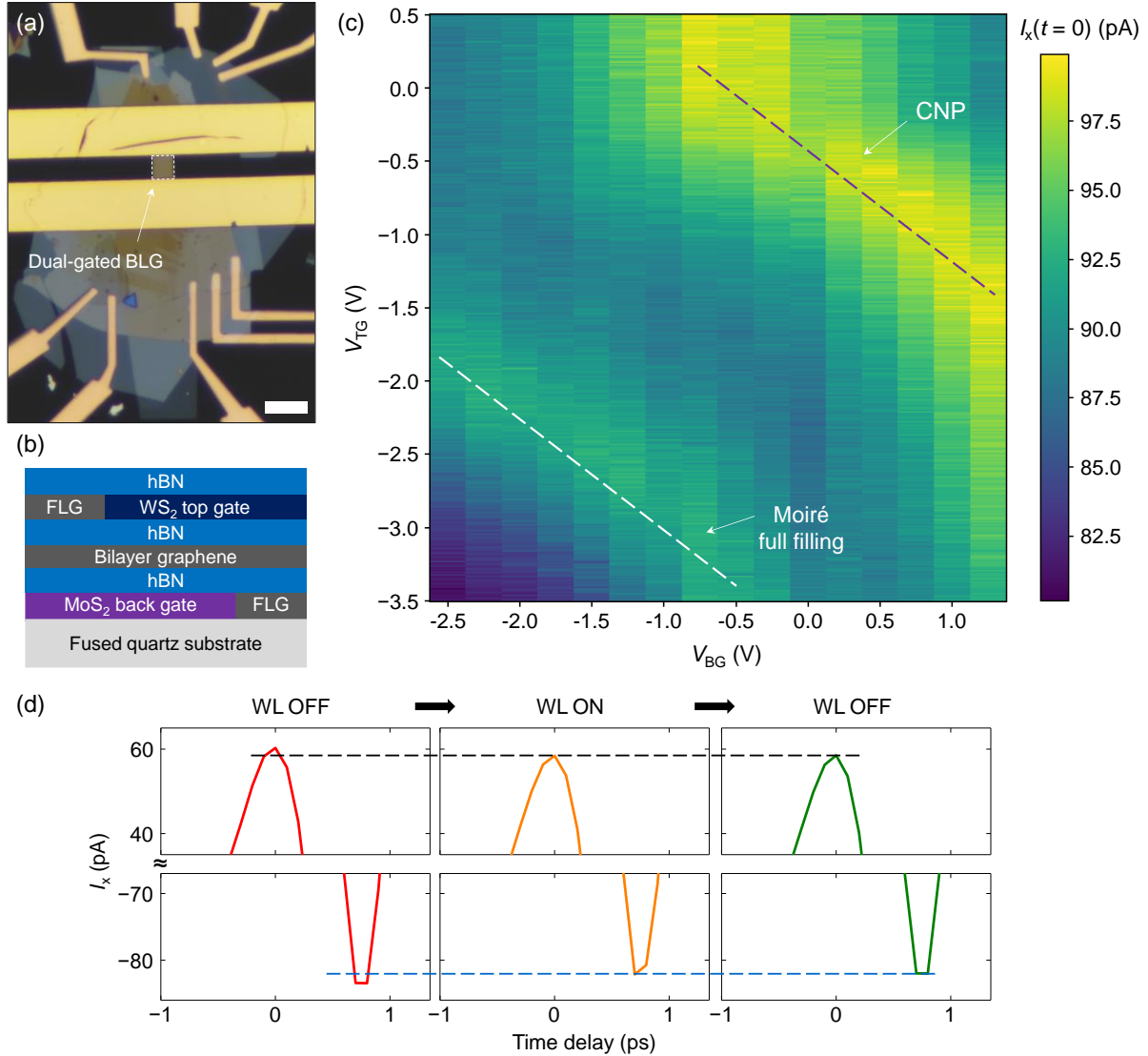

**Figure S5. Optical activation and deactivation of TMD gates in the device BLG3.** (a) Optical micrograph of the device BLG3. The thickness of bottom and top hBN dielectric was 25 and 18.5 nm, respectively. Scale bar, 15 μm. (b) Side-view schematic of the vdW heterostructure. (c) $I_x(t = 0)$ as a function of $V_{BG}$ and $V_{TG}$. The purple dotted line denotes a CNP line, and the white dotted line indicates a BLG/hBN moiré filling factor of −4. (d) Gate voltages were set to be $V_{BG} = 2$ V and $V_{TG} = -1$ V when the WL was turned off, and then the time-domain current $I_x$ was recorded (red). Afterwards, the WL was turned on and then $I_x$ was measured (yellow). Subsequently, $I_x$ was measured again after the WL was turned off (green). The maximum positive and negative current was same for the yellow and green curve (black and blue dotted line, respectively), which showed clear difference from those of the red curve.

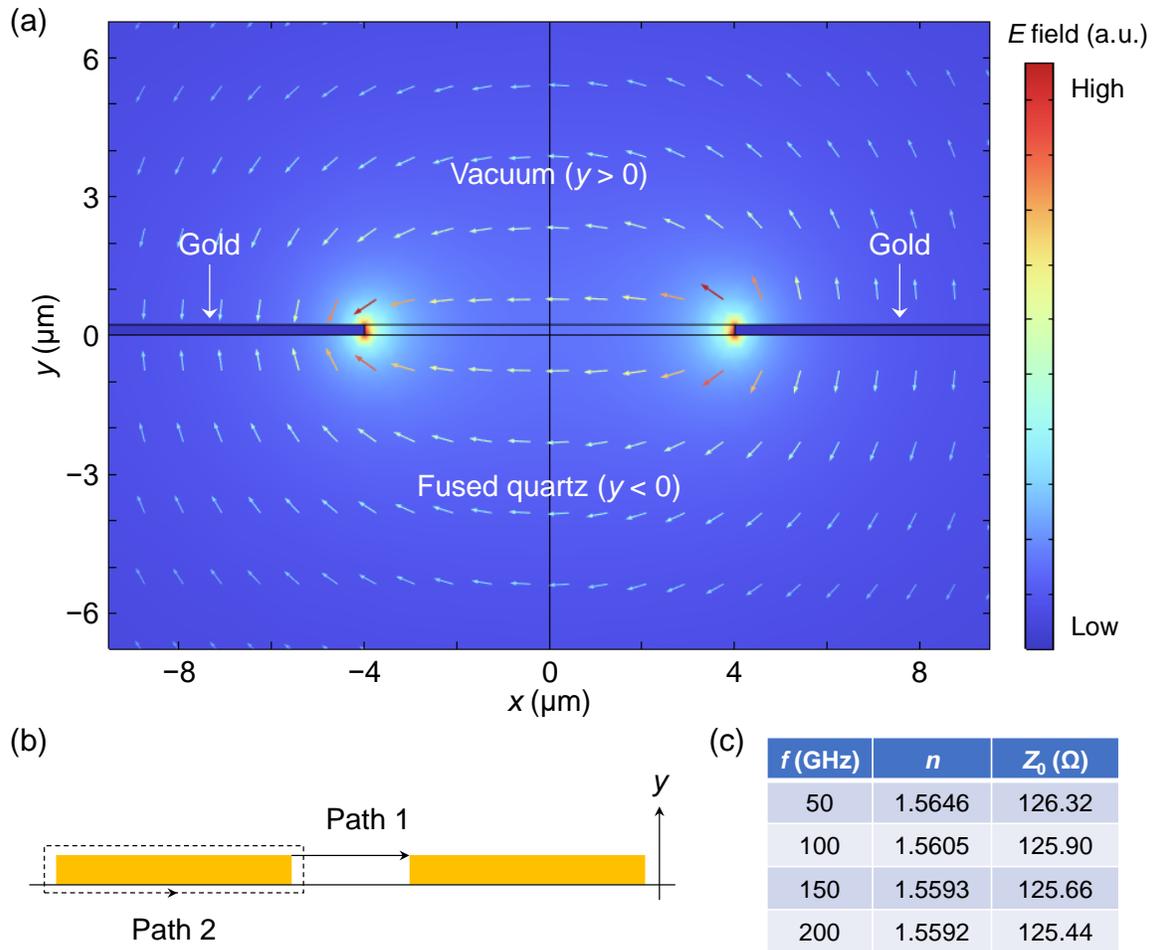

**Figure S6. Simulations of a hollow waveguide.** (**a**) Electric ($E$) field distribution of near a hollow waveguide at fixed frequency $f = 0.1$ THz. The half-plane $y < 0$ consists of fused quartz, and the other half-plane $y > 0$ is vacuum, where gold metal lines composing the waveguide is put at $y = 0$. (**b**) Schematics to describe how to calculate the characteristic impedance $Z_0$ of the waveguide. Path 1 was used to calculate potential difference $V$, and path 2 was used to calculate current $I$ flowing along the waveguide (yellow rectangles). (**c**) Refractive index $n$ and $Z_0$ calculated at several values of frequency $f$.

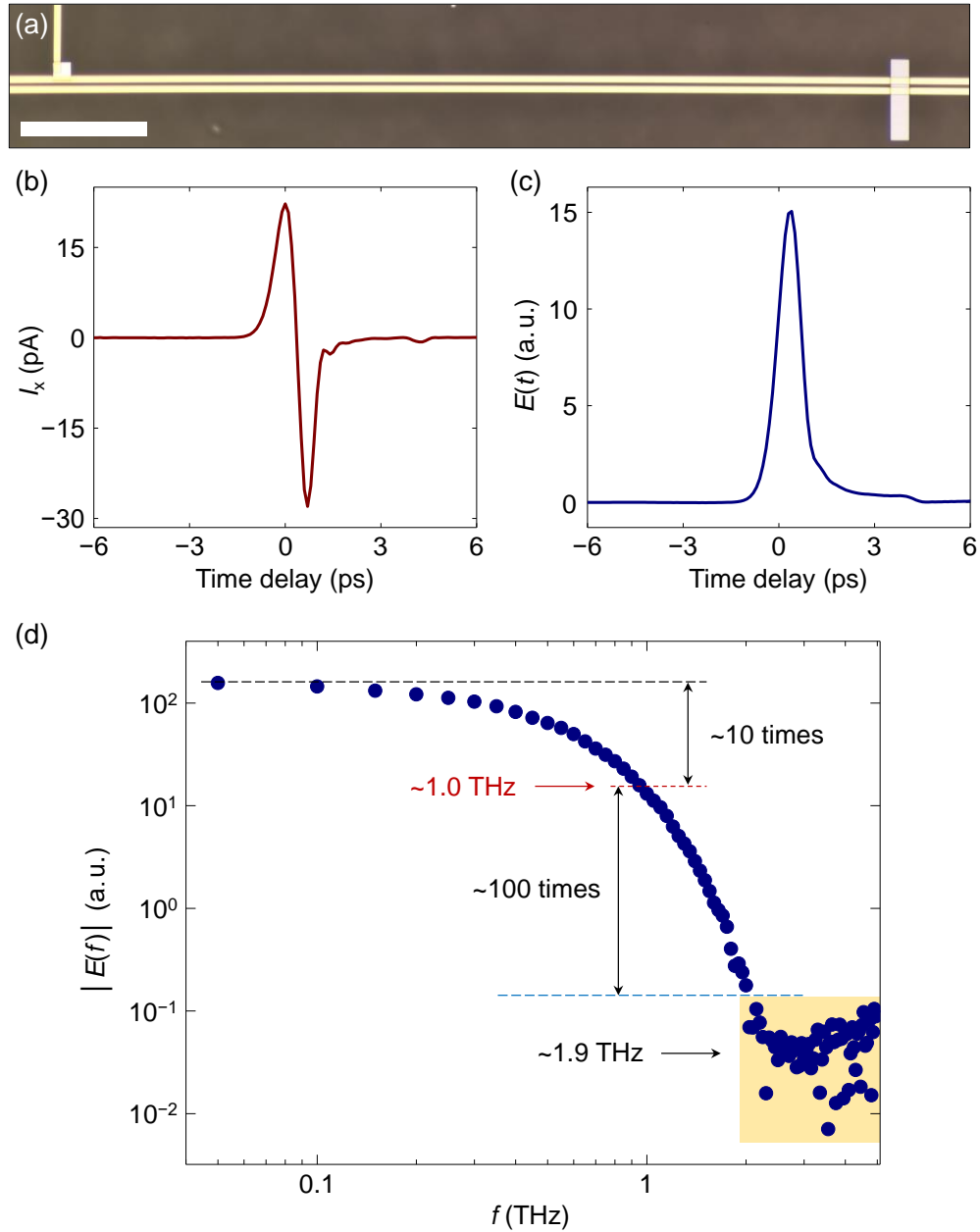

**Figure S7. Hollow waveguide and the bandwidth of our THz pulse.** (**a**) Optical micrograph of a hollow waveguide. Scale bar, 0.3 mm. (**b**) Time-domain current $I_x(t)$ measured at 4 K. (**c**) THz electric field $E(t)$ obtained from the numerical integration of data in (b). (**d**) Magnitude of $E(f)$, the fast Fourier transform of $E(t)$ in (c). The field intensity decreases from its maximum by ten times at $f \sim 1.0$ THz, and the noise floor (yellow rectangle) appears at $f \sim 1.9$ THz.

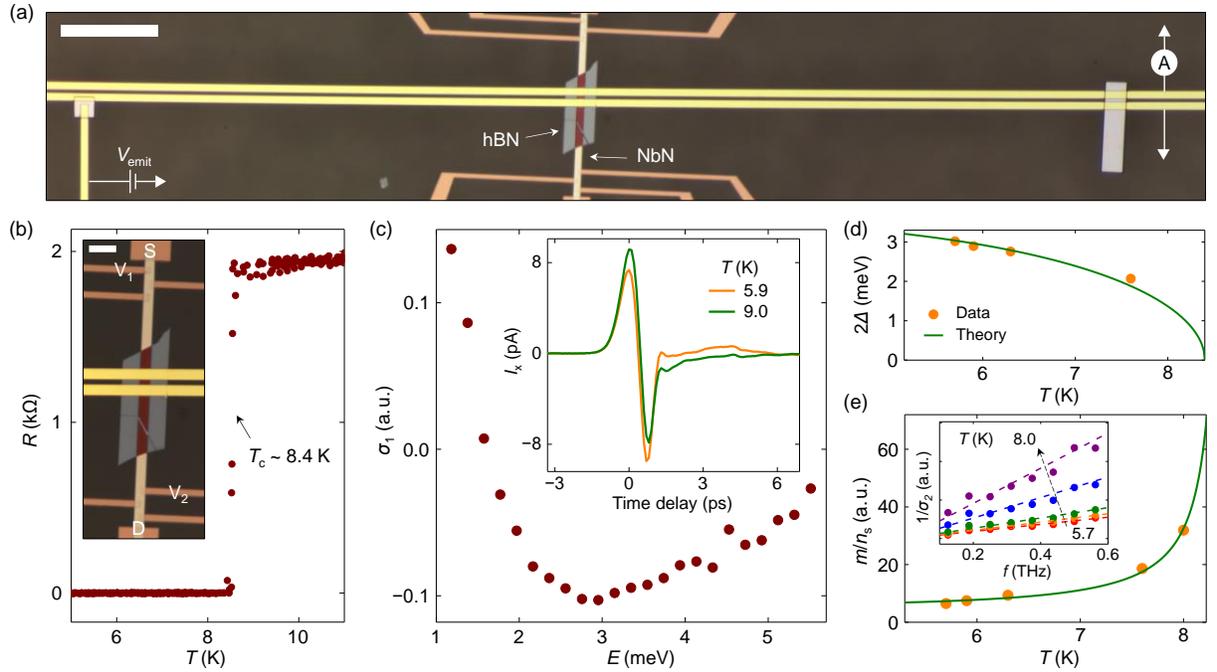

**Figure S8. THz spectroscopy on superconducting NbN.** (**a**) Optical micrograph of our NbN sample that is capped with hBN and embedded under THz waveguide. Scale bar, 0.2 mm. (**b**) DC resistance $R$ versus sample temperature $T$. NbN shows a clear superconductor transition at $T_c \sim 8.4$ K. Inset: Magnified image of the sample. Current was applied across the source (S) and drain (D), and voltage difference was measured between $V_1$ and $V_2$. Scale bar, 40 µm. (**c**) Real part of optical conductivity of NbN at $T = 5.9$ K. The spectrum shows a dip at ≈2.9 meV, which corresponds to the superconductor gap. Inset: Time-domain data measured at $T = 5.9$ and 9.0 K. (**d**) Superconductor gap measured at varied temperatures (yellow) and a BCS theory curve (green) fitted to the data. (**e**) Extracted superfluid stiffness at varied temperatures (yellow) and a phenomenological curve (green) fitted to the data. Inset: $1/\sigma_2$ versus frequency and linear fits at different temperatures.

| Material | Effective mass ($m_0$) | Reference | Additional details |
|---|---|---|---|
| $MoS_2$ | 0.56 / 0.64 | 3 | |
| $MoSe_2$ | 0.62 / 0.72 | 3 | |
| $WS_2$ | 0.33 / 0.43 | 3 | |
| $WSe_2$ | 0.35 / 0.46 | 3 | |
| Twisted bilayer graphene ($\theta = 1.05°$) | ≈0.05 | 4 | From $-4.5$ to $-3 \times 10^{12}$ cm$^{-2}$, approximately. |
| Twisted double bilayer graphene ($\theta = 1.33°$) | From ≈0.1 to ≈0.4 | 5 | From $-3$ to $-1 \times 10^{12}$ cm$^{-2}$, approximately. Constant $D$. |
| Twisted trilayer graphene ($\theta = 1.57°$) | From ≈0.5 to ≈1.0 | 6 | Constant $n \approx -2 \times 10^{12}$ cm$^{-2}$. $D/\varepsilon_0$ from $-0.8$ to $0$ V/nm. |

**Table S1. Effective mass *m* of TMDs and archetypal flat-band 2D systems.** For TMDs, the electron and hole mass are listed in order. For flat-band 2D systems, the range of a phase space in study is indicated in additional details.

| $n$ ($10^{12}$ cm$^{-2}$) | $D/\varepsilon_0$ (mV/nm) | $m$ ($m_0$) | $\tau$ (ps) | $\mu$ (m$^2$ V$^{-1}$ s$^{-1}$) |
|---|---|---|---|---|
| −1.2 | 60 | 0.0425 ± 0.0007 | 1.2757 ± 0.0356 | 5.279 ± 0.234 |
| −0.9 | 60 | 0.0403 ± 0.0006 | 1.3452 ± 0.0341 | 5.871 ± 0.236 |
| −0.6 | 60 | 0.0429 ± 0.0007 | 1.3826 ± 0.0390 | 5.668 ± 0.252 |
| −1.2 | 20 | 0.0450 ± 0.0007 | 1.4371 ± 0.0408 | 5.617 ± 0.247 |
| −0.9 | 20 | 0.0455 ± 0.0006 | 1.4579 ± 0.0343 | 5.636 ± 0.207 |
| −0.6 | 20 | 0.0480 ± 0.0013 | 1.4899 ± 0.0365 | 5.459 ± 0.282 |

**Table S2. Effective mass *m* and relaxation time $\tau$ extracted from the fitting processes of real optical conductivity $\sigma_1$ in Figure 4.** Errors originated from the fitting processes.